\documentclass[11pt]{article}
\usepackage{setspace}
\usepackage{amssymb}
\usepackage{amsmath}
\usepackage{amsfonts}
\usepackage{slashed}

\def\be{\begin{equation}}

\def\ee{\end{equation}}

\def\bea{\begin{eqnarray}}

\def\eea{\end{eqnarray}}

\makeatletter
\def\blfootnote{\xdef\@thefnmark{}\@footnotetext}
\makeatother

\begin{document}

\singlespace

\begin{flushright} BRX TH-6697 \\
CALT-TH 2022-001
\end{flushright}

\vspace*{.3in}

\begin{center}

{\Large\bf  A short pre-history of Quantum Gravity }

{\large S.\ Deser, FMRS \\}{\it 
Walter Burke Institute for Theoretical Physics, \\
California Institute of Technology, Pasadena, CA 91125; \\
Physics Department,  Brandeis University, Waltham, MA 02454 \\
{\tt deser@brandeis.edu}
}

\end{center}

\begin{abstract}

I describe the early, from the nineteen sixties, history of attempts at quantizing General Relativity.

\end{abstract}

The period 1915-1920 marks the appearance, and rapid acceptance, of Einstein's classical GR, one of the most revolutionary concepts in all of Physics, by one of the giants in quantum theory as well. Despite the war then raging in Europe, it was rapidly accepted by the elite of the time, though not as much by others until the verification of the ``three tests" in 1919, of which the most amazing was Einstein's calculation of the observed value of Mercury's perihelion precession, a non-linear effect. Then the field lay quiescent for a while because quantum theory and its successor, quantum mechanics, were so fertile at microscopic scales as to absorb all research for a long time.

It was unsurprising then, that about 15 years elapsed before any attempts at quantization were made, in the mid-thirties. These rather modest beginnings, motivated by early quantum field theory (QFT), consisted in quantizing the free massless spin 2 field, namely the linearized limit of GR in a weak field expansion about flat space. The two works are by Bronstein [1] (also Trotsky's real name) in the USSR and by Fierz and Pauli [2] in Switzerland. Pauli had a head-start: as a wunderkind aged 18, he wrote the first text on GR! Sadly, Bronstein who was also a wunderkind, was an early victim of Stalin's purges. Their quantization was a straightforward extension of that of electrodynamics, the other abelian gauge theory. All authors understood that this was far from a true QG, but at least it was its first glimmer, and indeed a long time --- about 25 years --- also passed until the next meaningful steps were taken. Remember that the idea of quantizing spacetime is conceptually vastly removed from that of ordinary matter systems, namely fermions and photons. This brings us to about 1960, when the reformulation of GR as a (rather unusual) field theory was successfully undertaken by Arnowitt, myself and Misner (ADM) [3] and similarly, if less completely, by Dirac [4]. Unlike theirs, most other GR work of the time involved its original classical geometrical form. I should also mention deWitt's lifetime devotion to this problem [5]. There was a separate interlude in the mid-fifties when several people (Klein, Landau, Pauli and I) independently suggested that QG might be a universal regulator for the infinities then ravaging matter loop calculations, but nothing came of it. My foolhardy attempt was the only published one [6], at the first conference devoted to GR at Chapel Hill, in 1957, where it was rapidly shot down by Feynman. [There seems also to have been a small meeting at the Bohr Institute in Summer 1957, at which I was apparently present, but of which I have no recollections whatever!]

Quite separately from this line is the lesson Heisenberg drew from Fermi's weak interaction model which, unlike quantum electrodynamics, had a dimensional coupling constant. He noted that any theory of this type would be beset by infinities of rising virulence with each perturbative order. This rapidly understood, if tacitly, insight applied to GR with Newton's constant, and cast a complete pall on QG. What Heisenberg told us was that perturbative (and there is no other way) QG was guaranteed to lose all predictive power as soon as one left its classical, tree, level. On the other hand, GR's quantization is mandated because Einstein's equations have matter as the source of gravity, so consistency requires it (I mercifully do not cite some feeble attempts to circumvent this by making the matter source be some sort of expectation value of its quantum nature). Even absent matter, GR must be quantized to avoid the ultraviolet catastrophe [7]. So the motivation to continue any QG program was greatly diminished. Still, ADM [8] and soon after, Schwinger [9] separately completed the formal quantization program. That is about the end of the prehistory, since there was little incentive to proceed further on this purely formal program, though one might add the decade later verifications of Heisenberg's predictions by explicit calculations of one-loop plus matter QG corrections, then still later, the two-loop ones of pure QG [10,11] in the seventies. A minor spinoff is the derivation of Classical GR using QM methods [12].

In summary, the prehistory of QG was a relatively short one --- consisting of formal quantization but without follow-through calculations of much interest, then explicit verification of non-renormalizability both of pure and matter-coupled GQ --- the mark of a field in search of motivation. How times have changed since then: observational cosmology has reignited the field, while Supergravity and String theory have become the chief theoretical QG arena, with mixed results so far, although a maximal version of SUGRA has survived finitely to at least seven loops, and of course superstrings are finite in the ten dimensions they require.  Even less successful have been some variants such as L(oop)QG, asymptotic safety and higher curvature models. Fortunately, I cannot discuss these within my early QG framework!

\section*{Acknowledgements}
I thank Prof. Alexander Blum for asking the many detailed questions that led to this essay, and useful correspondence.
This work was supported by the U.S.Department of Energy, Office of Science, Office of High Energy Physics under award number
de-sc0011632.

 \end{document}